# Impact of SiO$_2$ nanoparticle morphology on scattering efficiency for random lasers

Yan D. R. Machado[a], Gleice C. M. Germano[a],
Edison Pecoraro[b], Antonio Mario L. M. Costa[c] and Isabel C. S. Carvalho[a]

[a] *Physics Department, PUC-Rio - Pontifical Catholic University of Rio de Janeiro, Rio de Janeiro, 22451-900, Brazil*
[b] *Institute of Chemistry, UNESP - São Paulo State University, Araraquara, SP, 14800-060, Brazil*
[c] *Department of Chemical and Materials Engineering, PUC-Rio - Pontifical Catholic University of Rio de Janeiro, Rio de Janeiro, 22451-900, Brazil*

*Corresponding Author: Phone +55 21 995694282*
*E-mail address: yan.machado@live.com*

In this work, we compared the effect of Quartzite nanoparticles (QtzNP - crystalline SiO$_2$) and amorphous silica nanoparticles (ASNP) as scatters in a random laser (RL), correlating the laser efficiency to the morphology and refractive index (RI) of the scatterers. The RL was based on a laser media containing suspensions of QtzNPs or ASNPs in Rhodamine 6G (Rh6G) and ethylene glycol (EG) suspension. It was observed a considerable increase in RL efficiency for QtzNPs, as compared to ASNPs, even for concentrations as low as 0.5mg/mL. Such grow in the RL eficiency is attributed to the higher RI of QtzNPs (n=1.54), which increases the reflectance of light on the particles surface and the high field scattering intensity in the sharp edges of the flat slab shapes, associated to the particle's morphology. Finite Element Method (FEM) simulations were also performed for different particle size and shapes and an output increase is noticed for QtzNPs when compared with the ASNP confirming the experimental results. The comparison between the morphology of QtzNPs and ASNPs on scattering efficiency reported here contribute to optimize the performance of laser systems with natural and synthetic nanomaterials.

**Introduction**

Silicon-based optics have attracted a great deal of attention due to their increasing number of photonic applications in devices related to light generation, guiding, detection, amplification, and confinement [1,2].

A Random laser (RL) is a low coherence photon source in which the optical feedback relies on scattering processes in nano or submicron structured media, instead of a conventional mirror-based Fabry-Perot cavity. RL has attracted a great deal of attention in the last decades, as reviewed in [3], after its first clear experimental demonstration in 1994 [4], over two decades after Lethokhov's theoretical proposal of light being generated in a scattering medium with negative absorption [5]. The scattering media in the RLs are generally submicron or nanosized materials, either dielectric ($TiO_2$, $Al_2O_3$), semiconductors (ZnO), rare earth doped glasses and nanocrystals (Nd, Er) or metallic nanomaterials (e.g. Au, Ag), in which case localized surface plasmons plays a key role [6,7]. It is important to mention that random fiber lasers, whereby optical fibers are employed as the guiding media, have been demonstrated since 2007 [8], as reviewed in [3], and new varieties of guided-wave RLs are now being studied as a way to increase its range of applications.

Random lasers with nonresonant feedback based on Rh6G with sub-micron scatterers are widely studied. The quantum efficiency of the Rh6G solution depends on the dye concentration, while the emission efficiency of the RL is highly dependent on the transport mean free path (TMFP), which is related to the scatterers size, composition, concentration and morphology. Of all those characteristics that affect lasing efficiency, the influence of the morphology of $SiO_2$ nanoparticles as scatterers in liquid-solid suspension still lacks investigation. Similar work was done with $TiO_2$ nanoparticles (TNP) by Okamoto et al. [9] who studied how the morphology of TNPs affects the lasing efficiency in a dye-doped solid suspension (polymer) random media. Using TNP with a variety of shapes and sizes, they concluded that the peak emission intensity is highly dependent on particle morphology, while the spectral linewidth reduction is governed by change in the TMFP, which is dependent on the concentration of scatterers. Although they show the effect of the morphology of TNP on RL efficiency, they did not indicate which was the crystalline phase (CP) of those TNP (anatase, rutile or brookite), and did not compare the effect of the CP over RL efficiency. In crystals and anisotropic materials, the refractive index (RI) takes different values along the different axes, while in isotropic materials (amorphous structure), RI presents the same average values. Therefore, it is important to have both parameters accessed to weight the influence of each one on RL efficiency.

For a RL structure with cubic nanoparticles scatters, which exhibits some of the characteristics described in this work, Shuya Ning, et al. [10] showed a new configuration based on a plasmonic hybrid structure of (Au core)-($SiO_2$ shell) nanocubes deposited on Ag film (Au@SiO2 NC-Ag film), achieving a lower laser threshold when compared to spherical nanoparticles.

In order to gain insight based on experimental data, into how nanoparticle (NP) morphology affects efficiency in liquid fluorescent suspensions, which are the most common type of RL, we used silica ($SiO_2$), with two different structural phases and two different morphologies. Silica nanoparticles are widely employed as scatterers in the existing literature of RL because they present very good scattering properties and prevent photocatalytic degradation of fluorescent dyes [11,28], when compared to materials with higher refractive index, as $TiO_2$. Although amorphous silica nanoparticles with spherical or unspecified morphologies are commercially available or can easily be synthesized, nano or micro-sized crystalline $SiO_2$ particles (quartz) are comparatively costly and harder to obtain. Thus, in the present work, we choose to obtain the crystalline particles from a natural mineral source, *i.e.*, quartzite, which is low cost and widely available. Quartzite is a hard, non-foliated metamorphic rock formed from pure quartz sandstone, which is a classic sedimentary rock composed mainly of sand-sized (0.0625-2.00mm) quartz-silicate grains. This sandstone is converted into quartzite rocks through extreme heat and pressure, usually related to tectonic processes like compression within orogenic belts. Quartzite rocks are made of aggregates of micro and nano quartz slab shapes, which shows higher refractive index (n=1.54) than amorphous $SiO_2$ (n=1.46) for the entire wavelength visible range [12]. Obtaining those NP from quartzite avoids the usual complexes methods used to prepare silicon-based materials for RLs [14], which requires meticulous preparation, several chemical reagents and some expensive equipment. The production of QtzNPs does not require any chemical setup besides mechanical grinding and the material's characterization to determine the range of size and particle's morphology.

We systematically studied the way morphology of scatters influences lasing performance with natural, low-cost quartzite micro and nanocrystals and further confirm results with finite element method simulations using COMSOL Multiphysics.

The simulations further confirm that QtzNP morphology (namely the sharp edges of its flat slabs) improves the scattering for the small particles (Mie Scattering Regime) with the far scattered field showing intensities when compared to the incident light. Meanwhile, silica nanoparticles with no such morphology (spherical in the case of the simulation), present a scattered field with shorter range than the rectangular-square shapes of the QZNPs, making it impossible for RL action to appear for lower concentrations of scatters (0.5mg/mL) and with high efficiency when compared to other published work, as can be seen in the survey below.

**TABLE 01 - Survey of Random Lasers with similar oxide-based scatterers**

| Scatter | Structure | RI$_{scatter}$ | Solvent | RI$_{solvent}$ | Gain Medium | Concentration Scatters | Treshold Energy | FWHM (nm) | Ref |
|---|---|---|---|---|---|---|---|---|---|
| Quartzite (SiO$_2$) | Crystalline | 1.54 | EG | 1.42 | Rh6G 1mM | 0.1 - 15mg/mL * | 0.25mJ 0.568mJ/cm * | ~6nm | This Work |
| Au@SiO$_2$ | Cubic particles with SiO2 Shell | - | Poluestirene | 1.58 | Alq3 + DCJTB | - | 6.7 – 31.1 uJ / cm² * | 1.1-12.8nm * | [10] |
| SiO$_2$ | Spherical nanopartilce | 1.47 | Ethanol | 1.36 | Rh640 0.1mM | 10^12 – 3 * 10^13 particles/Ml * | 20 - 100mJ/cm² * | 7.5 – 17.5 * | [11] |
| SiO$_2$ | Doped Amorphous Particles | 1.16 | Ethanol | 1.36 | Rh6G 6.6mM | Doped particles | 24uJ / pulse RL Pulse FWHM 100ps | | [13] |
| SiO$_2$ | Porous Porous Monolith | - | Ethanol | 1.36 | Rh6G 0.1mM | - | 9.7uJ/pulse | 4nm | [15] |
| TiO$_2$ | PMMA Polymer Doped film | 2.61 | Ethanol | 1.36 | Rh6G 0.001mM – 1mM | 10^-3 mol/L | 15mJ | 9nm | [16] |
| TiO$_2$ | Aggregate, Needle, Rod, Spindle, Rice, Bow-tie, Butterfly | 2.50 – 2.70 * | Adell K40 Polymer | 1.50 | Rh6G 5.0mM | 5 – 20% volume fraction of particles * | - | 3.5-5.5nm * | [9] |
| TiO$_2$ | Rutile- spherical nanoparticle | 2.61 | Methanol | 1.32 | Rh6G 1mM | 3.54 – 30.0 mg/mL * | 12.48mJ/cm² | 5-11nm * | [17] |
| TiO$_2$ | Doped Polymer with nanowire | 2.61 | Poly(N-vinylcarvazole) | 1.68 | Rh6G | - | 0.44mJ | <0.8nm | [18] |
| ZnO | Spherical nanoparticle Film | 2.00 | - | - | ZnO powder | 5.6g/cm³ | 763kW/cm² | 0.3nm | [19] |
| TiO$_2$ @ Silica | Amorphous Particles with SiO2 Shell | - | Ethanol | 1.36 | Rh6G 0.1mM | 5.6*10^10 particles/mL | 15mJ/cm² | 5nm | [20] |
| TiO$_2$ @ Silica | Amorphous Particles with SiO2 Shell | 2.61 | Ethanol | 1.36 | Rh6G 1mM | 5.6*10^10 particles/mL | 2.28mJ | ~5nm | [21] |
| TiO$_2$ @ Silica | Spherical Core-shell | 2.61 | Ethanol | 1.36 | Rh6G 0.1mM | 140*10^10 particles/mL | 1.2mJ/cm² | ~7nm | [22] |

*depending on sample

**Materials and Methods**

The QtzNP was produced by grinding quartzite rocks (acquired from a local store) in a high-hardness agate mortar for several minutes. Grinding the material with a conventional ceramic mortar would result in impurities due to quartzite's hardness. The amorphous silica nanoparticles (Column Chromatography Silica Gel QingdaoH >98%) (ASNP) undergone the same process.

Random laser solutions with Rhodamine6G (Sigma-Aldrich 95%) in ethanol (Sigma-Aldrich>99.9%) and methanol (Sigma-Aldrich 99.7%) were tested and worked, but both were not viscous enough to result in a stable suspension of the particles, and in a few minutes, a volume of precipitated particles was notice on the bottom of the container. To overcome this problem, we choose ethyleneglycol (Sigma-Aldrich >99%) (EG) as solvent. It results in a slightly lower quantum efficiency when compared to ethanol, which influences the threshold energy, since the refractive index of EG (1.43) is higher than that of methanol and ethanol (1.36) [23,11]. However, its viscosity is high enough to yield a stable suspension for both quartzite and silica nano and microparticles found in the present work. A higher energy threshold for the suspension using EG as solvent is expected [24].

Materials composed by quartz aggregates offer a higher refractive index than amorphous silica particles for every wavelength in the visible region [12], and it is known that the difference between the gain media and scatterer's refractive indexes, has influence in the random laser efficiency [3]. The higher RI of QtzNP is due to its crystalline structure, which results in higher surface reflectance for visible light that further justifies the good results for nano and micro-sized scatterers.

**SEM Characterization of Quartzite**

In order to confirm the average size and morphology of QtzNP and ASNP, the RL suspensions were deposited onto aluminum stubs, dried, and coated by sputtering with thin gold layer (Q150T ES Quorum sputter coater). Figure 1 shows has images that were acquired with a TESCAN Clara (UHR-FEM) electron microscope. ImageJ software (version 1.45s) was used to infer the particles sizes, which range from 50nm to 500nm, with an average value of 225nm for QtzNP and 100nm for ASNP. Energy dispersive X-ray spectroscopy (EDS) microanalysis shows that the ASNP and QTZNP are composed of 72.39% O, 27.61% Si, without the presence of impurities like hematite, zirconia or titania.

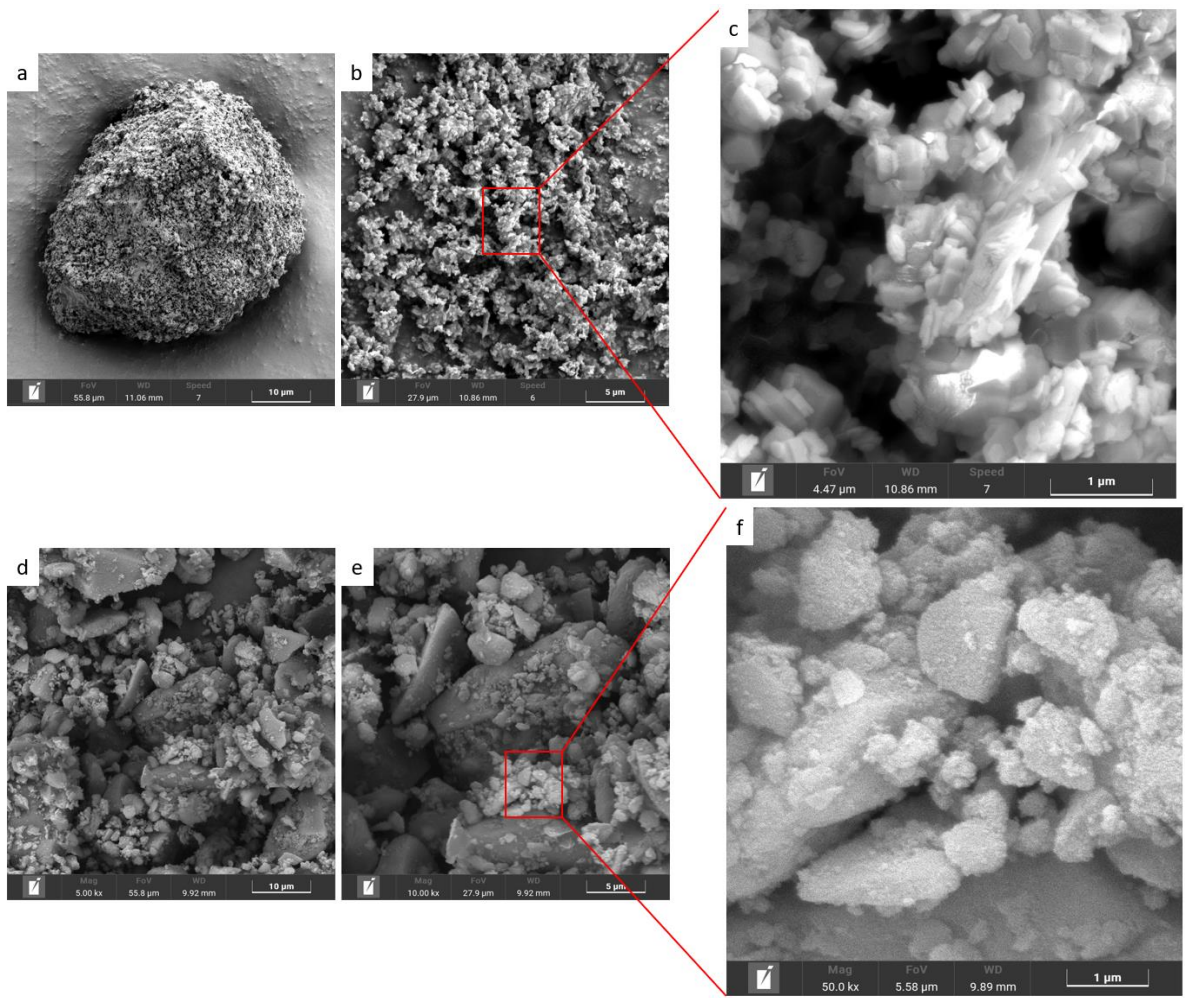

*Fig. 1: Images obtained by SEM-FEG of (a-c) quartzite as agglomerates and nanocrystals and (d-f) amorphous silica nanoparticles.*

**Random Laser Characterization**

The RL emission spectra of the Rhodamine-6G (Rh6G)/QtzNP and Rh6G/ASNP suspensions were measured upon excitation with a pulsed Q-Switched Nd:YAG laser (Brio-Quantel, 10 Hertz, 6ns, wavelength = 532nm). The pump beam carries energies that can vary up to 10s of mJs with a ~2.1mm beam diameter. The beam passes through both, a dichroic mirror and a bandpass filter (BG18 Thorlabs - 412-569 nm), to prevent infrared light from degrading the sample, while also having its energy controlled with a half-waveplate (Thorlabs GL15-A) and a polarizer mounted on a circular rotatory mount. The spectral analysis was performed with an Ocean Optics USB4000-UV-VIS spectrometer (spectral resolution of 1.5nm), and the measurements were all done within 15 minutes at room temperature. Figure 2 shows the experimental setup and a picture of the suspension in a glass.

The excitation light was directed into the sample at a 45-degree angle to avoid creating a Fabry-Pérot cavity or whispering gallery modes due to reflections on the walls of the 4mL quartz vial.

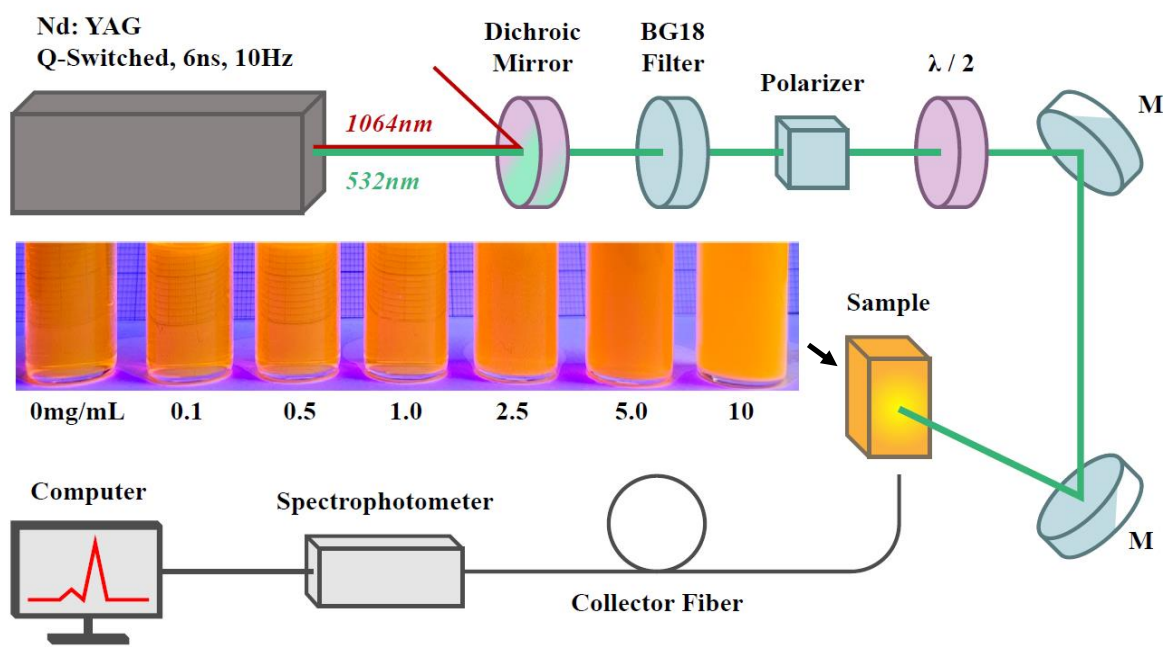

*Fig. 2*: RL Experimental setup for the suspensions in the quartz cuvette. The inset shows images of the samples for different QtzNps concentrations in EG/Rh6G (1mM).

**Results on Random Laser Behavior**

*Rh6G + Quartzite Nanoparticles suspension*

The laser emission from the sample is evidenced by a significant (~30nm) spectral narrowing under increasing excitation regimes (Fig 3). Furthermore, Figure 4 displays the exponential growth in the fluorescence intensity as a function of the excitation energy, resulting in a well-defined laser threshold excitation energy for which the gains from feedback in the fluorescent media are greater than losses due to absorption and light scattering.

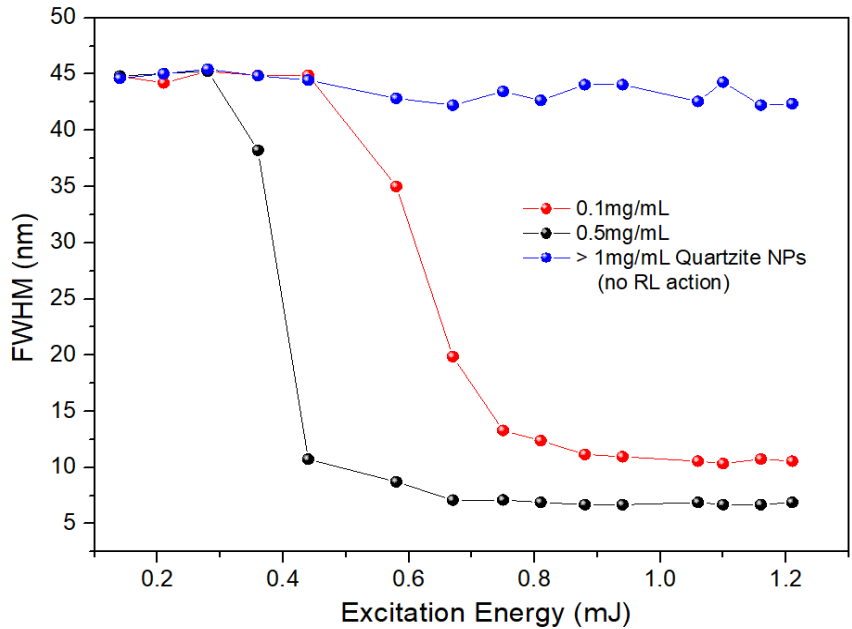

*Fig. 3*: FWHM of Rh6G emission as a function of the excitation energy for different QtzNP concentrations in the suspension.

Figure 5 shows the emission intensity as a log plot in order to analyze the response of the random laser for different QtzNP concentrations and determine the RL threshold for each suspension. The optimal QtzNP concentration found is 0.5mg/mL, with a threshold of 0.25mJ. Therefore, it is possible to conclude that increasing the concentration further, after this value, the RL threshold increases, reducing the laser efficiency. This fact can be attributed to the light scattering mean free path, which is too small for higher concentrations making the interaction between scattered light and the fluorescent dye molecules too short before new scattering events, thus not providing enough gain to surpass the losses due to absorption effects. Tests conducted for concentrations below 0.5mg/mL also showed reduction of the lasing intensity and even not lasing at all, even with very high excitation energies. This behavior is due to scattering events happening too far from each other to provide the necessary feedback.

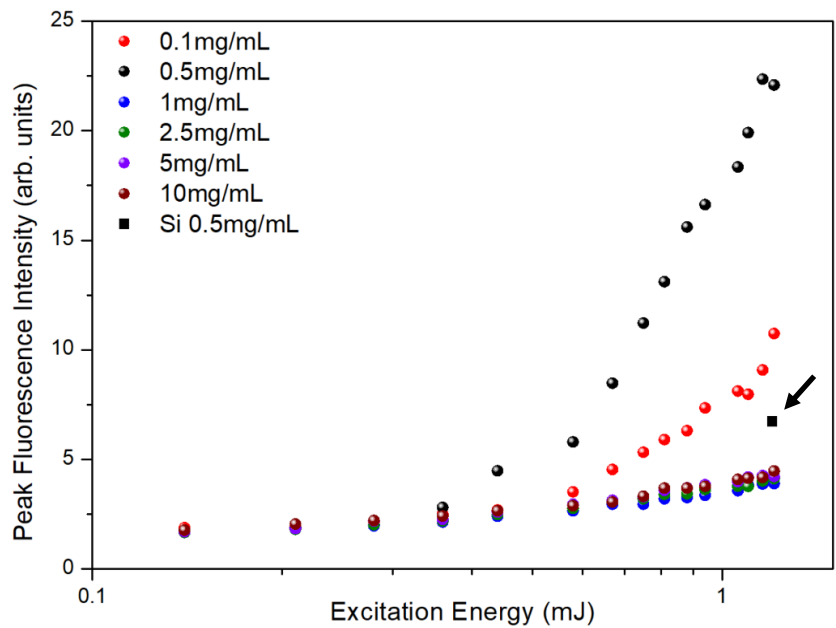

*Fig. 4*: RL intensity as a function of the excitation energy for various QtzNP concentrations.

Figure 5 shows the comparison between results of RL emission for QtzNP and ASNP samples, with 0.5mg/mL concentration of scatterers for both suspensions. At the same excitation energy (~2mJ, well above the threshold) it is clear that QtzNP offer a higher efficiency due to a more noticeable increase in both, the fluorescence intensity (Fig. 5a) and reduction of the FHWM (Fig. 5b).

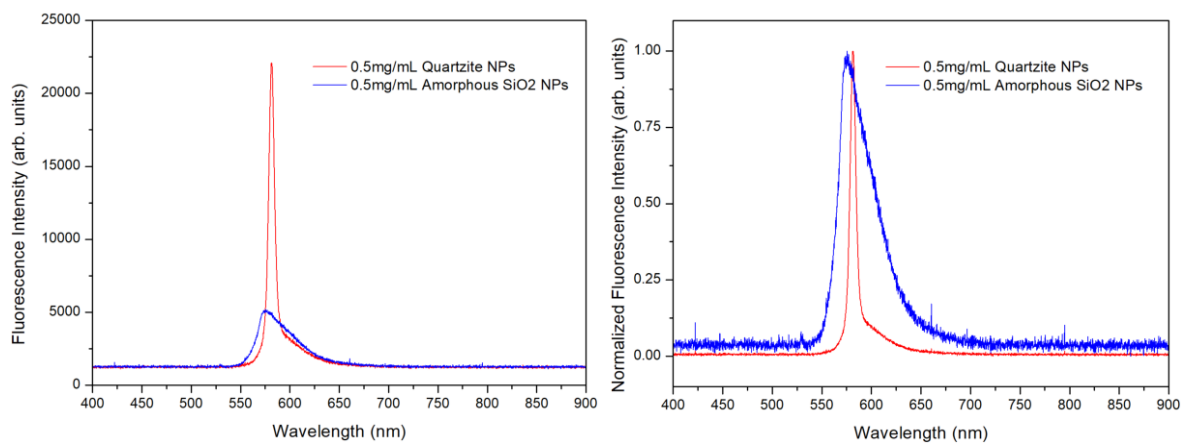

*Fig. 5*: RL emission comparison for QtzNP and ASNP suspension with the same scatterers concentration.: (a) fluorescence intensity as a function of the wavelength, showing higher intensity emission for QtzNP; (b) normalized emission showing a higher reduction in the FHWM for the sample containing QtzNP.

In order to analyze the emission spectra in more detail, fluorescence peak wavelength was plotted as a function of the QtzNp concentration. In Figure 7 it is clear that the peak emission wavelength increases with the concentration of scatterers.

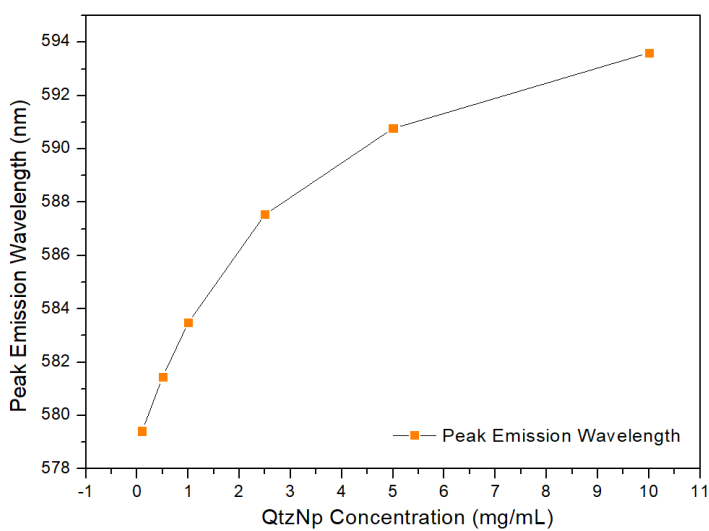

*Fig. 6*: RL peak emission wavelength as a function of concentration of QtzNp scatterers

The increase in peak emission wavelength indicates that the dye molecules are isolated for lower concentration of QtzNp, having value closer to that of pure Rh6G + EG as the spectra is dominated by the monomer fluorescence. As the concentration of QtzNp increases, we observe a consistent shift towards higher peak wavelength. This is a well-known phenomenon that happens in Rh6G solutions for high dye concentrations [24], typical of dimerization of the dye molecules. The process leads to a redshift in the sample's fluorescence to the spectral region dominated by these aggregates' emission. However, the same behavior was observed here, correlated to the scatterers concentration.

We suggest that it could arise from two processes: first, an increase in the reabsorption of the fluorescence photons by the Rh6G molecules, due to the decrease in the TMFP associated to the increase in scatterers density; second, an increase in the occurrence to form dimmers on the surface of the scatterers. This assisted dimerization could be explained by the increasing in the total $SiO_2$ surface area (due to the increase of the NP concentration), on which Rh6G molecules could be trapped by intermolecular interactions with the $SiO_2$ substrate, increasing the probability to interact by similar mechanism with others Rh6G molecules, resulting in the formation of dimmers. This change in fluorescence peak wavelength also justifies the lower efficiency of higher QtzNp concentration samples (Fig. 4), which can be related either to dimerization processes, or to the intersection in the absorbance and emission spectra, allowing for less reabsorption of light for excitation energies below threshold [24].

**Finite Element Method – Light Scattering Simulations & Discussion**

In order to understand if the role played by the scatterers nanoparticles morphology is responsible by the results above, we investigate the electromagnetic scattering for nanometer-scale objects with computational simulations. We used COMSOL Multiphysics software to model our systems through the finite element method (FEM) [25] using the Electromagnetic Waves, Frequency Domain (ewfd) interface.

Figure 7a shows the simulation results for nanoparticles with the same crystalline phase (assigned through the same value for refractive index), but with different morphology, *i.e.*, spherical and prismatic, both with 75nm radius. The scattering for perfectly spherical particles with sizes comparable to the wavelength of the incident light can be precisely calculated using Lorentz-Mie and Rayleigh solutions, depending on the size of the particle, and are well documented in literature [26].

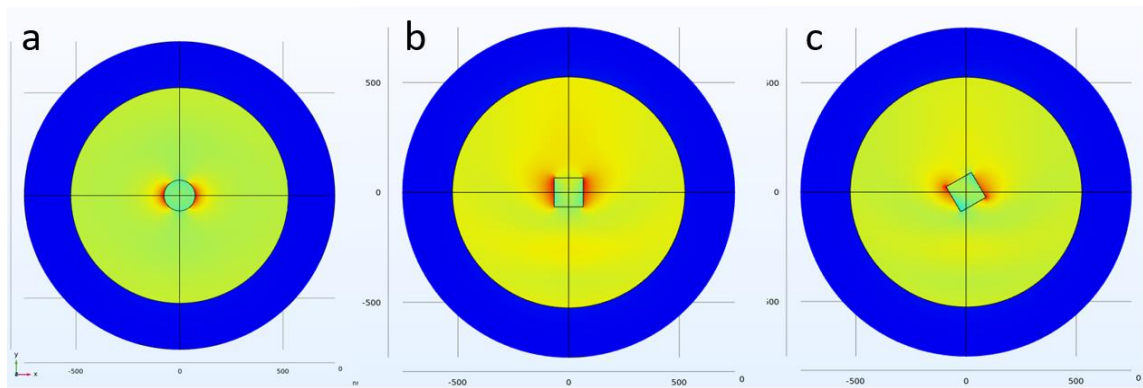

*Fig. 7*: Normalized scattering patterns of the electric field for different NPs morphologies. The source electric field is defined to reach the particle surface at z+ axis and is scattered according to Mie's solutions, and the scattered light intensity is higher in the perpendicular direction of the incident light.

The QtzNP material consisting of crystalline aggregates was defined by the refractive index and the morphology to simulate simple morphologies with the characteristic sharp edges and high reflectance faces of the nanocrystals (Fig. 7b 7c). With this approach we can verify the scattering patterns and why QtzNP offers high efficiency as scatterers. For that, we simulated two geometries, while preserving the scattering cross section used for the nanosphere and the aspect ratio of particles found in the microscopy measurements.

It is important to notice the concentration of field is present at the faces of the prism perpendicular to the incident field, aligned to the polarization, as expected. These regions, where the field is considerably higher, are important in order to provide a sufficient gain for the fluorescent dye to surpass absorption losses, thus increasing the efficiency of the scattering process for the random laser with particles presenting sharp symmetrical edges.

The source field used was defined at a maximum value of 10V/m for all simulations and it is possible to notice that, when compared to the conventionally used spherical silica particle, prismatic morphology, as those in QtzNP, can offer regions with higher scattered field values and lower loss for longer distances. This reinforces the hypothesis that the sharp edges and high reflectance faces of the nano-crystalline material are key factors in increasing random laser efficiency, and also justifies our material being able to offer a higher scattering mean free path, thus making lasing possible with lower scatterers concentrations, as observed in our experimental configuration.

It is important to mention that the higher refractive index of quartzite, when compared to silica nanoparticles [12], is another factor that further improves the efficiency of the scattering process [3]. Nonetheless, it was not simulated because our background material is defined as air and not the fluorescent dye – since we are only interested in visualizing how light interacts with materials that present new morphologies. More than that, our material not only scatters light (for smaller particles), but also reflects it (for the micro and nano sized particles and aggregates that remain from the grinding process), making it more probable to form localization phenomena in the gain medium, increasing the efficiency of the RL even more [27].

**Conclusion**

Morphology of silica nanoparticles directly influences RL efficiency, and, through experiments and simulations, it is noticeable that crystalline aggregates (QTZNP) with flat faces and sharp edges are more efficient in scattering light than amorphous (ASNP) or spherical nanoparticles, since they increase the scattered field range and feedback. When comparing our samples laser threshold with the other RLs reported in literature, with similar scatterer composition but different morphologies, a lower threshold energy is observed in survey 1. Finite element method simulations confirm that in the scattering process the efficiency is higher for QTZNP due to the crystalline morphology of the sample.

**Acknowledgments**

This work was partially supported by the Brazilian agencies FINEP, CAPES, FACEPE, CNPq and Instituto Nacional de Fotônica – INCT/INFo. Y.D.R.M., I.C.S.C. and A.S.L.G. acknowledge support by the Office of Naval Research Global (ONRG). The authors thank Edison Pecoraro for the quartzite material and all fruitful discussions. Yan D. R. Machado acknowledges Mr. Fredy for technical support and designer F. Werneck for helping with the experimental setup figure. The authors are grateful to CBPF for the use of the facilities for electron microscopy in the LabNano/CBPF, Rio de Janeiro-BR (JEOL JSM-7100FT) and GReo/PUC-Rio for microscopy services and the use of TESCAN/Clara electron microscopy.

**Disclosures**

The authors declare no conflicts of interest. Parts of this work were presented at the Latin America Optics and Photonics (LAOP) Conference 2022 and published as an abstract in the Technical Digest Series (OPTICA Publishing Group, 2022) as cited [28].